\documentstyle[aps,twocolumn]{revtex}
\begin{document}
\title{The reliability horizon for semi-classical quantum gravity:\\
Metric fluctuations are often more important than back-reaction}
\author{Matt Visser\cite{e-mail}}
\address{Physics Department, Washington University, St. Louis, 
         Missouri 63130-4899}
\date{gr-qc/9702041; 20 February 1997; Revised 18 June 1997}
\maketitle
\section*{Abstract}
{\small

In this note I introduce the notion of the {\em ``reliability
horizon''} for semi-classical quantum gravity. This reliability
horizon is an attempt to quantify the extent to which we should
trust semi-classical quantum gravity, and to get a better handle
on just where the {\em ``Planck regime''} resides. I point out that
the key obstruction to pushing semi-classical quantum gravity into
the Planck regime is often the existence of large metric fluctuations,
rather than a large back-reaction. There are many situations where
the metric fluctuations become large long before the back-reaction
is significant. Issues of this type are fundamental to any attempt
at proving Hawking's chronology protection conjecture from first
principles, since I shall prove that the onset of chronology
violation is always hidden behind the reliability horizon.

{\em Revisions:} Central definitions and results essentially
unchanged. Discussion of the relationship between this letter and the
Kay--Radzikowski--Wald singularity theorems greatly extended and
clarified. Discussion of reliability horizon near curvature
singularities modified. Several references added. Minor typos
fixed. Technical \TeX\ modifications.

}
\section{Introduction}

Semi-classical quantum
gravity~\cite{Centenary,Birrell-Davies,Fulling,Visser} is the
approximation wherein we keep the gravitational field classical,
while quantizing everything else. This hybrid theory is clearly a
half-way-house on the road to quantum gravity, but it has two very
decided advantages over the various speculative theories of quantum
gravity currently in vogue: (1) Semi-classical quantum gravity is
firmly based in experimental reality, and (2) Semi-classical quantum
gravity is relatively tractable. In particular, essentially all of
black hole thermodynamics, the stress-energy renormalization
programme~\cite{Birrell-Davies,Fulling}, and Lorentzian wormhole
physics~\cite{Visser,Morris-Thorne,MTY,Visser89a,Visser89b},  is
carried out within this framework.

A related issue that is currently only treatable within the framework
of semi-classical quantum gravity, and that has excited much recent
attention, is that of Hawking's {\em chronology protection
conjecture}~\cite{HawkingI,HawkingII}. Various attempts at proving
this conjecture have been
made~\cite{Visser,Kim-Thorne,Frolov91,Visser93,Visser94}, culminating
in the singularity theorems of Kay, Radzikowski, and Wald~\cite{KRW}.
These theorems are set up, proved, and in fact only make sense
within the semi-classical quantum gravity approximation, so it
behooves us to develop a rather good feel for where we should stop
believing this approximation.  Qualitatively, the answer to this
question has been known since the pioneering work of Wheeler in
the late 1950's~\cite{Wheeler55,Wheeler57}: We should certainly
stop believing semi-classical quantum gravity once we enter the
Planck regime.  The subtleties arise in recognizing the onset of
Planck scale physics.

(Of course there is the logical possibility, however unlikely, that
all current theories could break down with only just a little more
energy in our accelerators.  I will not deal with any new physics
that may be hiding just around the corner, but will instead focus
on the region where we know that current theories must break down of
their own accord even if no other new physics intervenes at lower
energies.)

To help get a better handle on these issues I will introduce a few
new concepts: the {\em ``reliable region''}; the {\em ``reliability
boundary''}; and the {\em ``reliability horizon''}. These concepts
will be defined in a way that is similar to standard concepts of
classical general relativity: the chronology-violating region;
chronology boundary; and chronology horizon respectively.  The
precise location of reliability horizon I define is (deliberately)
somewhat fuzzy, in many respects it is similar to the stretched
horizon of the black hole Membrane Paradigm~\cite{Membrane}. We
can always argue over the last factor of ten or so concerning where
exactly to place the reliability horizon.  This fuzziness is not
a problem and the notion of reliability horizon is still useful
since I shall show that the reliability horizon is always outside
the chronology horizon.

Physically, the reliability horizon is characterized by the onset
of {\em either} large metric fluctuations (Planck scale curvature
fluctuations) {\em or} a large back-reaction (Planck scale expectation
value for the curvature). I shall argue that in many situations
the onset of large metric fluctuations precedes the onset of large
back-reaction. In view of this  we should stop believing the
semi-classical quantum gravity approximation at the reliability
horizon, {\em before} reaching the chronology horizon.

\section{The chronology horizon}

Let $\gamma$ be any geodesic (spacelike, null, or timelike) that
connects some point $x$ to itself. Let $\sigma_\gamma(x,y)$ denote
the relativistic interval from $x$ to $y$ along the geodesic $\gamma$.

\begin{equation}
\sigma_\gamma(x,y) = 
\left\{
\matrix{+s^2               &    \hbox{if the geodesic is spacelike,}\cr
        \hphantom{+}0      &    \hbox{if the geodesic is lightlike,}\cr
        -\tau^2            &    \hbox{if the geodesic is timelike.} \cr}
\right.
\end{equation}

Define level sets $\Omega(\ell^2)$ by

\begin{equation}
\Omega(\ell^2) \equiv 
\left\{x:\exists \gamma\neq0 | \sigma_\gamma(x,x) \leq \ell^2\right\}.
\end{equation}

If the metric were Riemannian, rather than Lorentzian, these level
sets would be very simple. In particular we would have
$\Omega(0)=\emptyset$. Because the metric is indefinite the set
$\Omega(0)$ has the possibility of being nontrivial. Note that it
is essential that the geodesic in question not be the trivial
geodesic from $x$ to itself, otherwise $\Omega(\ell^2)$ would
trivially be the whole spacetime for $\ell^2 \geq 0$.

The set $\Omega(0)$ is called the {\em ``chronology violating
region''}.  By definition any $x\in\Omega(0)$ is connected to itself
by a nontrivial timelike or null geodesic---as such we clearly have
the possibility of time travel from the point $x$ to itself.

The set ${\cal B} \equiv \partial[\Omega(0)]$ is the {\em ``chronology
boundary''}---this is the boundary that we will have to cross in
order to actively participate in time travel effects.

The set ${\cal H}^+ \equiv \partial [J^+(\Omega(0))]$ is the {\em
``chronology horizon''}---it is the boundary of the future of the
chronology violating region.  This is the boundary that we will
have to cross in order to passively participate in time travel
effects. (Passive participation is the ability to see time travel
effects somewhere in one's past, without being able to influence
the past. This is quite sufficient to thoroughly disrupt known
physics and is just as reprehensible as active participation.)

Note that these definitions only make sense if we are dealing with
a fixed Lorentzian geometry---other interesting quantum field
theoretic processes may be going on, but the geometry is fixed and
unquantized. This is exactly the statement that we are dealing with
semi-classical quantum  gravity.  These definitions are fundamental
to analyses of the chronology protection conjecture: The basic idea
is that something must go wrong on or near the chronology horizon.

\section{Chronology Protection Theorems}

Working entirely within the context of semi-classical quantum
gravity, Kay, Radzikowski, and Wald~\cite{KRW} showed that
in any spacetime containing a compactly generated chronology horizon
there must exist certain points on the chronology horizon such that
the two-point function is not locally of Hadamard form. (Similar
results hold for a broad class of non-compactly generated chronology
horizons.)

This is to be contrasted with the observations of
Sushkov~\cite{Sushkov,Sushkov-2},
Krasnikov~\cite{Krasnikov,Krasnikov-2}, and the present
author~\cite{Roman-ring}. For instance, Sushkov~\cite{Sushkov} showed
that in Misner space, a particularly simple exemplar for a spacetime
with chronology horizon [not compactly generated], a specific choice
of quantum field theory coupled with a specific choice of quantum
state keeps the stress-energy regular all the way to the chronology
horizon. In addition, Krasnikov~\cite{Krasnikov} exhibited several
$(1+1)$--dimensional geometries with bounded stress-energy near the
chronology horizon (Cauchy horizon). Finally, the present author has
shown that with enough wormholes it is possible to keep the stress
energy arbitrarily mild arbitrarily near the chronology
horizon~\cite{Roman-ring}.

The Sushkov and Krasnikov results can be brought into conformity with
the Kay--Radzikowski--Wald (KRW) results by noting that the KRW
results show only that the stress-energy is ``singular'' at some
points on the chronology horizon---and that the word singular in this
context can either mean infinite or may mean simply mean
undefined~\cite{KRW,Kay-Cramer}.

Additional investigations bearing on this matter are those of
Boulware~\cite{Boulware}, Grant~\cite{Grant}, and Tanaka and
Hiscock~\cite{Tanaka-Hiscock}.

The KRW results are formulated as stress-energy singularity theorems
in semi-classical quantum gravity without back-reaction. It is only
{\em after} we add back-reaction (by insisting that the spacetime
satisfy the semi-classical Einstein equations) that we begin to
see the outlines of a chronology protection theorem: Any spacetime
that satisfies the semi-classical Einstein equations must at the
very least have all two-point functions locally of Hadamard form
(otherwise we cannot even define the stress-energy tensor) and
therefore cannot contain a compactly generated chronology horizon.

Stated differently, if we try to insert a chronology horizon into an
otherwise respectable spacetime satisfying the semi-classical Einstein
equations then either (1) some points on the chronology horizon are
curvature singularities, or (2) the spacetime no longer has a uniquely
defined geometry (because it no longer has a uniquely defined
stress-energy tensor) and so is no longer even a manifold in the
normal sense.

The key point to focus on here is that the Kay--Radzikowski--Wald
singularities occur at the actual onset of chronology violation---at
the chronology boundary. By a {\em reductio ad absurdum}
argument~\cite{Private}, this may be interpreted as indicating the
breakdown of semi-classical quantum gravity at the chronology
horizon. Since by this argument the chronology horizon cannot be
treated by semi-classical methods it follows that time machines
themselves cannot be globally treated by semi-classical methods:
Time machines (if they exist at all) intrinsically require the use
of full quantum gravity to describe the chronology horizon.  Time
machines (if they exist at all) are thus intrinsically ``beyond
the pale'' of known physics.  This is an important and crucial
result: it indicates that the {\em apparently plausible} manipulations
typically invoked to turn a traversable wormhole into a time machine
are much less plausible under closer scrutiny.

The KRW argument, while definitively indicating the breakdown of
semi-classical quantum gravity at (at least some points on) the
chronology horizon, unfortunately gives very limited information
as to what if anything goes wrong near the chronology horizon. I
shall now argue that it is possible to introduce the notion of a
reliability horizon to qualitatively describe the breakdown of
semi-classical quantum gravity and shall argue that we should not
trust the semi-classical approximation to quantum gravity beyond
this reliability horizon. In particular, and in agreement with this
interpretation of the KRW results, we should not trust the
semi-classical approximation to quantum gravity at the chronology
horizon.

\section{The reliability horizon}

I will define (and justify the definition of) the reliability
horizon in three stages, refining the definition as we develop
additional insight.

\medskip

{\bf Definition 1a:} Using the notation developed above, let
${\cal U} \equiv \Omega(+\ell_{Planck}^2)$ be the {\em ``unreliable
region''}. It consists of those points $x$ that are connected to
themselves by spacelike geodesics as short as, or shorter than,
one Planck length. 

(It is here that the fuzziness of the reliability horizon is made
manifest: If you wish to argue for a safety margin by requiring
the spacelike geodesics to be longer than, say, ten Planck lengths,
very few people would want to argue with you. Note further that we
should not blindly declare all Planck-length geodesics to be
meaningless. The existence of photons, which in the geometrical
optics limit are treated as particles moving along null geodesics,
shows that at least some geodesics of length zero make perfectly
good sense in semi-classical gravity. It is only topologically
nontrivial closed geodesics that are captured by the above
definition---see the physical justification for this definition
below.)

The entire thrust of this definition is that it is an attempt to
give an invariant and unambiguous meaning to the notion ``within a
Planck length of the chronology horizon'', an invariant interpretation
of this phrase being necessary before it is possible to decide
where the Planck regime resides.

\medskip

{\bf Definition 1b:} Let ${\cal B}_{Planck} \equiv \partial [{\cal
U}] \equiv \partial [\Omega(+\ell_{Planck}^2)]$ be the {\em
``reliability boundary''}---this is the boundary that we will have
to cross in order to actively probe the unreliable region.

\medskip

{\bf Definition 1c:} The set ${\cal H}^+_{Planck} \equiv
\partial[J^+({\cal U})] \equiv \partial [ J^+(\Omega(+\ell_{Planck}^2))]$
is the {\em ``reliability horizon''}---it is the boundary of the
future of the unreliable region.  This is the boundary that we will
have to cross in order to passively probe the unreliable
region~\cite{Alternatives}.

\medskip

{\bf Justification:} At this stage these are merely definitions,
they do not carry any weight until we physically justify the
terminology. The physics behind these definitions is that I shall
show, extrapolating from low energies, that at the very least the
unreliable region as defined above will be subject to large
fluctuations in the metric even if the expectation value of the
curvature is relatively mild~\cite{Kuo-Ford}.

(It is in addition possible that current theories break down by
the introduction of new physics long before we get to the Planck
scale. I am not addressing that possibility but am only interested
in the unavoidable breakdown of semi-classical quantum gravity
assuming that no new physics intervenes.)

To see some more technical details of the physics behind these
definitions it may be helpful to consider the set

\begin{equation}
\Delta\Omega = \Omega(+\ell_{Planck}^2) - \Omega(0).
\end{equation}

\noindent
This set consists of all points that are connected to themselves
by nontrivial ultrashort Planck length spacelike geodesics, but
which do not themselves suffer from the additional complication of
being in the chronology violating region. Any quantum field
that we try to set up in this region $\Delta\Omega$ is automatically
subject to Planck scale physics.

By itself, this is not enough to justify the terminology ``unreliable'',
but it is enough to encourage a deeper look at the issues. Continuing
our preliminary analysis: If we were to think of trying to do a
mode sum for the quantum field (quasi-Fourier decomposition) then
the fact that in at least one direction the field has a periodicity
less than $\ell_{Planck}$ implies that the mode sum will contain
momenta of order $E_{Planck}$ and higher.  Two cases of extremely
high symmetry can be used to clarify this point: Consider a
(3+1)--dimensional hyper-cylinder of circumference $\ell <
\ell_{Planck}$. Any quantum field defined on such a hyper-cylinder
can be decomposed into an infinite tower of (2+1)--dimensional
quantum fields.  These will consist of a low-energy mode coming
from the translationally invariant mode plus an infinite stack of
particles more massive than the Planck mass.  The same sort of
thing happens with a spherically symmetric traversable wormhole of
throat radius less than one Planck length; the mode decomposition
now runs over spherical harmonics (plus an undetermined radial
mode). Again we obtain an infinite stack of trans-Planckian particles,
this time (1+1)--dimensional particles labelled by the angular
momentum quantum numbers. This is certainly enough to justify calling
$\Omega(+\ell_{Planck}^2)$ the {\em ``trans-Planckian region''}.

Now as long as all the quantum field theories in question are
renormalizable we should not worry about this trans-Planckian
physics: after all the key aspect of renormalizable theories is
that we do not need to know the details of the high energy features
of the theory---renormalizable theories are precisely those for
which we can still make low energy predictions without worrying
about ultra-high energy phenomena.

{\em The key point, however, is that Einstein gravity is itself
known to be non-renormalizable}~\cite{Wheeler55,Wheeler57}. If we
had a renormalizable theory of quantum gravity then the entire
discussion of the reliability horizon would be moot. The reliability
horizon has to do with the breakdown of our trust in the semi-classical
theory based on Einstein gravity---it is a consequence of the
non-renormalizability of Einstein gravity and not a fundamental
limitation on the as yet undiscovered theory of full quantum gravity.

Indeed, if we take Einstein gravity and linearize it about the
background we are interested in, we can then ask how the linearized
gravitons behave as quantum fields  on this background geometry.
The resulting quantum field theory is well known to be non-renormalizable
with a dimensionful coupling constant given by the Planck
mass~\cite{Wheeler55,Wheeler57}. Once we enter the unreliable region
these linearized gravitons are subject to Planck scale physics
which in this  non-renormalizable theory is definitely a disaster.
Inside the unreliable region the linearized gravitons will be
strongly interacting (and also unitarity violating) and will thereby
lead to Planck scale fluctuations in the curvature, even if the
expectation value of the curvature is pleasingly mild.

It is ultimately these large metric fluctuations and associated
Planck scale curvature fluctuations that tell us that we should no
longer trust semi-classical quantum gravity behind the reliability
boundary. This is finally enough to justify calling
$\Omega(+\ell_{Planck}^2)$ the {\em ``unreliable region''}.

\medskip

{\bf Definition 2:} An improvement of the previous definition, if
the manifold in question is multiply connected one, is to keep
track of the winding number of the geodesic. (In spacetimes containing
traversable wormholes this will just be the number of times the
geodesic threads through one of the wormholes.) Decompose the
homotopy classes of self-intersecting geodesics emanating from the
point $x$ into equivalence classes $\Gamma_N$ characterized by
winding number $N$, and define
 
\begin{equation}
\Omega_N(\ell^2) \equiv 
\left\{x:\exists \gamma\in \Gamma_N| 
\sigma_\gamma(x,x) \leq N^2\ell^2\right\}.
\end{equation}

\noindent
The point is that if a geodesic wraps through $N$ wormholes and is
of length less than $N\ell$, then at least one
wormhole-to-wormhole segment of the curve must be of length less
than $\ell$. Now simply replace $\Omega(\ell^2)$ by

\begin{equation}
\Omega_\infty(\ell^2) \equiv 
\Omega(\ell^2) \cup (\cup_{N=1}^\infty \Omega_N(\ell^2))
\end{equation}

\noindent
in all definitions regarding the reliability region.

\medskip

{\bf Definition 3:} The definition given above still does not
capture all of the situations in which we should cease trusting
semi-classical quantum gravity.  We should also not trust regions
where the background manifold exhibits Planck scale curvature. We
shall consider the sets

\begin{equation}
\Omega_R(\ell^2) = 
\left\{x: R_{\mu\nu\sigma\rho} R^{\mu\nu\sigma\rho} > \ell^{-4} \right\}.
\end{equation}

\noindent
Thus we shall characterize curvature singularities only by their
scalar invariants.

(Unfortunately, as is well known form the theory of spacetime
singularities, simply looking at the curvature invariants is not
really sufficient to characterize even curvature singularities
\cite{Wald}. The problem lies in the fact that the Lorentzian metric
is indefinite so that the curvature invariant can be zero even if the
curvature tensor is not zero. This is a standard problem in the
characterization of curvature singularities whose resolution requires
technically messy complications beyond the scope of this letter.)

\noindent
With the above definition $\Omega_R(0)$ is the set of
curvature-invariant singularities of the spacetime, and
$\Omega_R(\ell_P^2)$ is a set of points for which at least one
component of the Riemann tensor takes on Planck scale values.

We now augment the unreliable region by setting

\begin{equation}
{\cal U} \equiv  \Omega_\infty(+\ell_{Planck}^2) 
\cup \Omega_R(+\ell_{Planck}^2).
\end{equation}

\noindent
Similarly, the reliability boundary becomes

\begin{eqnarray}
{\cal B}_{Planck} &\equiv& \partial[{\cal U}],
\nonumber\\
&\equiv&
\partial \left[ \Omega_\infty(+\ell_{Planck}^2)
\cup \Omega_R(+\ell_{Planck}^2) \right],
\end{eqnarray}

\noindent
and the reliability horizon becomes

\begin{eqnarray}
{\cal H}^+_{Planck} &\equiv& \partial[J^+({\cal U})], 
\nonumber\\
&\equiv& 
\partial[J^+\left(\Omega_\infty(+\ell_{Planck}^2) 
\cup \Omega_R(+\ell_{Planck}^2) \right)].
\end{eqnarray}

\section{Are the chronology protection theorems physically reliable?}

The application to the chronology protection theorems is immediate:
If you interpret the KRW singularity theorems as chronology protection
theorems (via the {\em reductio ad absurdum argument}~\cite{Private})
then you deduce the breakdown of semi-classical quantum gravity at
the chronology horizon.

Since the chronology horizon is always, by definition, inside the
reliability horizon I have just defined, I would in addition argue
that we should never trust the physical applicability of semi-classical
quantum gravity at or near the chronology horizon. Despite the fact
that the expectation value of the stress energy can be made
arbitrarily small at and near the chronology
horizon~\cite{Sushkov,Sushkov-2,Krasnikov,Krasnikov-2,Roman-ring},
the analysis of this note argues that metric fluctuations will
become Planck scale once one crosses the reliability horizon.

To prove a {\em physically trustworthy} version of chronology
protection, we would need either: (1) a theorem within the context
of semi-classical quantum gravity that makes reference only to
physics outside the reliability horizon, or (2) a theorem within
the context of full-fledged quantum gravity.

Unfortunately, an acceptable theory of full-fledged quantum gravity
does not yet exist, and (as mentioned above) for the approximate
hybrid theory called semi-classical quantum gravity there are known
to be many classes of geometries for which back reaction is negligible
all the way up to the reliability
horizon~\cite{Sushkov,Krasnikov,Roman-ring}, in fact we can keep
the back reaction negligible up to the chronology horizon. (In
these situations it is the onset of large fluctuations in the metric
which characterizes the reliability horizon.) The fact that any
such geometry exists shows that option (1) above is impossible,
and leaves us contemplating all the complications of full quantum
gravity itself.

This is compatible with the viewpoint~\cite{Private} wherein one
views the KRW singularity theorems as providing a {\em reductio ad
absurdum} disproof of the reliability of semi-classical quantum
gravity at the chronology horizon. The new features here are that
the reliability horizon gives us qualitative control over where
this breakdown occurs and that I have exhibited plausible physical
mechanisms behind this breakdown.

This does {\em not} imply that we have succeeded in completely
circumventing chronology protection---I do not mean to imply that
this analysis shows that we can actually succeed in building a time
machine.

What the this letter does is to drive home the point that semi-classical
quantum gravity is fundamentally incapable of answering the question
of whether or not the  universe is chronology protected.  To really
answer this question we will first need an acceptable theory of
quantum gravity. (My own point of view on this topic has shifted
markedly over the last few years---I was initially very hopeful
that chronology protection could be proved from first principles
within the context of semi-classical quantum gravity.)

Given the impossibility of proving chronology protection
within the framework of semi-classical quantum gravity, and the
concomitant need to address quantum gravity itself, maybe it might
be a good idea to adopt chronology protection as an axiom---and
let it help guide us to an acceptable theory of quantum
gravity~\cite{Visser}.

\acknowledgements

This research was supported by the U.S. Department of Energy. 

I wish to thank Bernard Kay and Robert Wald for their thoughtful
comments.



\begin{references}
\bibitem{Alternatives}
Earlier attempts at developing notions somewhat analogous to the
reliability horizon can be found in~\cite{HawkingI,HawkingII}
and~\cite{Kim-Thorne}. These early suggestions involve an undesirable
observer dependence~\cite{Visser}.
\end{references}
\end{document}